\documentclass[12pt,twoside,a4paper]{article}
\usepackage[dvips]{epsfig}
\def\gsim{\:\raisebox{-0.5ex}{$\stackrel{\textstyle>}{\sim}$}\:}
\def\lsim{\:\raisebox{-0.5ex}{$\stackrel{\textstyle<}{\sim}$}\:}
\begin{document}
\thispagestyle{empty} 
\title{
\vskip-3cm
{\baselineskip14pt
\centerline{\normalsize DESY 00--044 \hfill ISSN 0418--9833}
\centerline{\normalsize MZ-TH/00--09 \hfill} 
\centerline{\normalsize hep--ph/0003148\hfill} 
\centerline{\normalsize March 2000 \hfill}} 
\vskip2cm
Gluon Fragmentation to Gluonium  \\[3ex]
\author{H.~Spiesberger$^1$ and P.~M.~Zerwas$^2$ \\[2ex]
\normalsize{$^1$ Institut f\"ur Physik,
  Johannes-Gutenberg-Universit\"at,}\\ 
\normalsize{Staudinger Weg 7, D-55099 Mainz, Germany} \\[0.7ex]
{\normalsize $^2$ Deutsches Elektronen-Synchrotron DESY, D-22603
  Hamburg, Germany}\\[13ex]
} }
\date{}
\maketitle
\begin{abstract}
\medskip
\noindent
The fragmentation of gluons to gluonium states is analyzed qualitatively
in the non-perturbative region. The convolution of this mechanism with
perturbative gluon radiation leaves us with a hard component in the
fragmentation of gluon to gluonium.
\end{abstract}

\newpage 

Theoretical analyses of gluonic matter particles were initiated
\cite{fritzsch} soon after gluon fields were introduced as the basic
force fields of the strong interactions. Mass spectra and quantum
numbers of such states have been studied in several approaches to
non-perturbative QCD \cite{chan}. Since gluon-rich environments of
collision processes are the preferential source for the production of
gluonia, several mechanisms of this kind have been analyzed at great
detail, in particular heavy quarkonium decays \cite{roy+walsh} and
Pomeron processes \cite{close}. Strong candidates have been observed in
various experimental analyses \cite{landua}, though final conclusions
could not be drawn yet.

Recently it has been suggested to search for gluonium states in $Z$
decays \cite{roy}. A significant fraction of these decays involves gluon
jets so that the LEP/SLC $Z$ events provide a natural ensemble to search
for such states. The energy of these gluon jets is large enough to
justify a parton picture for the production of gluonia. This process can
be described by the convolution of the hard cross section for producing
a gluon with a fragmentation function accounting for the transition of
gluons to gluonia.  The resulting expressions are correct up to terms
which decrease as inverse powers of the gluon-jet energy.

In this note we describe an attempt to predict the momentum spectrum of
the gluonium particles within the fragmented gluon jets. The solution of
this problem is of experimental relevance in optimizing search
strategies for gluonia. In particular, if a major fraction of the gluon
energy is transferred to the gluonium states in the fragmentation
process, the experimental identification is facilitated {\em vis-a-vis}
the soft fusion of gluons to gluonia in the plateau region for which the
gluonium decay products submerge to the low-energy hadron sea.

\noindent
\underline{Non-perturbative Fragmentation:} The basic idea for the
primordial non-perturbative fragmentation of gluons to generic gluonium
states\footnote{To analyze the gross features of the fragmentation
  function, it is sufficient to consider $gg$ gluonium states as
  example.} $G$ of mass $M_G \simeq 1.5$ GeV follows the path of
heavy-quark fragmentation, the gross structure of which can be described
by the Peterson et al.\ fragmentation function \cite{peterson}. A simple
form of the fragmentation function can be derived by adopting the
quantum-mechanical rules of old-fashioned perturbation theory
\cite{nishijima} to estimate transition probabilities in the parton
model. The qualitative features of the amplitude for a high-energy gluon
$g$ to fragment into a gluonium state $G$ with a fraction $z$ of the
gluon momentum are determined by the energy transfer $\Delta E = E_G +
E^{\prime}_g - E_g$ across \linebreak
\begin{figure}[hb]
\unitlength 1mm
\vspace{-1cm}
\begin{picture}(120,33)
\put(20,-3){\epsfig{file=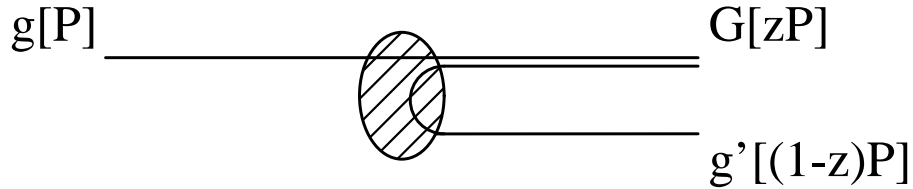,,width=12cm}}
\end{picture}
\caption{\it The non-perturbative $g \rightarrow G$ fragmentation in the 
  parton picture.}
\label{fig1}
\end{figure}

\noindent
the vertex in the fragmentation process (see Fig.\ \ref{fig1})
which conserves three-momentum:
\begin{equation}
{\rm Amplitude} \left[g \rightarrow G + g^{\prime} \right] \propto
\Delta E^{-1} \, . 
\label{amplitude}
\end{equation}
Expanding the energies for large gluon momenta $P$ about the gluonium
mass $M_G$ and the transverse momentum which can be assumed of order of
the strong interaction scale $\Lambda$,
\begin{equation}
\begin{array}{rcl}
\Delta E &= &\sqrt{M_G^2 + z^2 P^2} 
             + \sqrt{\Lambda^2 + (1-z)^2 P^2} - P 
\\[1ex]
         &\propto 
            & \displaystyle 
              \frac{1}{z} 
            + \frac{\epsilon_G}{1-z} \, ,
\end{array}
\label{deltaE}
\end{equation}
and taking into account the standard factor $z^{-1}$ for the
longitudinal phase space which generates the non-per\-tur\-bative
rapidity plateau, we suggest the following ansatz for the
non-perturbative $g \rightarrow G$ fragmentation function:
\begin{equation}
\displaystyle
d_g^G(z) = \frac{N}{z \left(\frac{1}{z} + \frac{\epsilon_G}{1-z}
  \right)^2} \, .
\label{dgg}
\end{equation}
The shape parameter $\epsilon_G$ is defined as 
\begin{equation}
\epsilon_G = \Lambda^2/M_G^2
\end{equation}
according to the expansion Eq.\ (\ref{deltaE}). 

\begin{figure}[t]
\unitlength 1mm
\begin{picture}(120,90)
\put(20,0){\epsfig{file=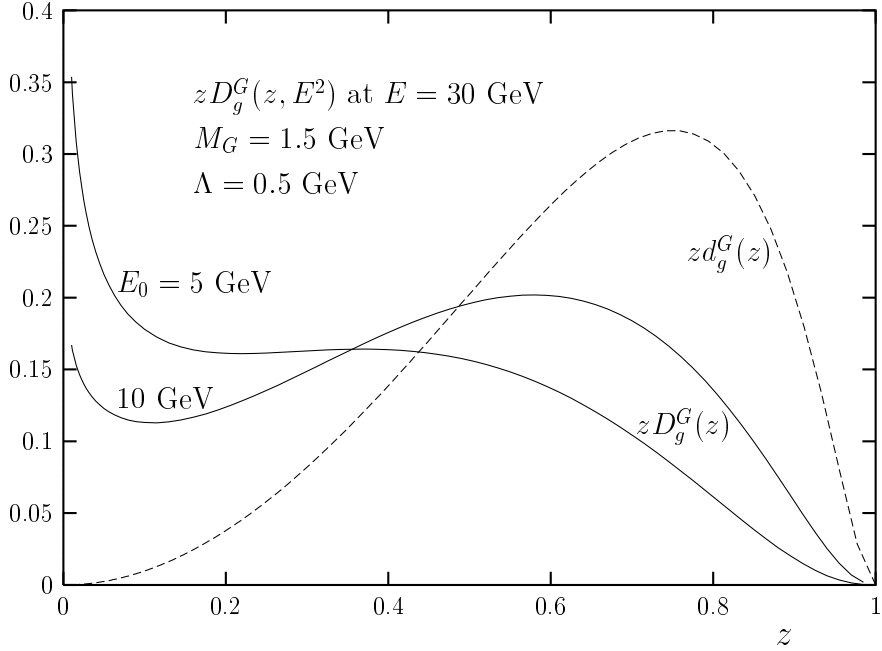,,width=12cm}}
\end{picture}
\caption{\it Gluon-to-gluonium fragmentation at the energy $E = 30$
  GeV. Dotted line: non-perturbative fragmentation function $d_g^G(z)$
  with $N = 1$ (for the sake of clarity the fragmentation functions are
  multiplied with the momentum fraction $z$); full lines: fragmentation
  function $D_g^G(z,E^2)$ after the evolution from the scales $E_0 = 5$
  and $10$ GeV to the energy $E = 30$ GeV.}
\label{fig2}
\end{figure}

The coefficient $N$ denotes the normalization which after integration
over the momentum spectrum defines the (unknown) rate at which gluons
fragment non-perturbatively into gluonium states. Choosing for
illustration $M_G = 1.5$ GeV and $\Lambda = 0.5$ GeV, the small
coefficient $\epsilon_G \simeq O(1/10)$ makes $d_g^G(z)$ peak strongly
near $z = 1$ (dashed curve in Fig.\ \ref{fig2}). The maximum of the
momentum distribution is predicted at $z_{\rm max} \simeq 1 -
\sqrt{2\epsilon_G}$ for small $\epsilon_G$. This form is a
straightforward consequence of quantum mechanics which enhances
transitions with minimum energy transfer. For large momenta, i.e.\ large
$z$, the impact of the heavy gluonium mass on the energy transfer is
less effective than for small momenta.

This fragmentation picture applies only for large energies of the
fragmenting gluon for which $\Delta E \propto P^{-1}$ approaches zero. A
lower bound of the required gluon energy may be estimated by demanding
the energy transfer $\Delta E$ to be less than a fraction of the typical
strong interaction scale $\Lambda$. It follows from the inequality
$\Delta E \lsim \Lambda/2$ that the gluon energy must exceed
\begin{equation}
E \gsim E_0 \simeq M_G^2/\Lambda
\label{E0}
\end{equation}
in the laboratory frame which amounts to about 5 GeV for the parameters
introduced above. 

\noindent
\underline{Perturbative Gluon Radiation:} If the gluon energy exceeds
the minimum value (\ref{E0}), the parent gluon is attenuated by
secondary gluon bremsstrahlung at early times before the
non-perturbative fragmentation becomes operative. These perturbative QCD
effects can be described in analogy to the Altarelli-Parisi evolution
\cite{ap}. Neglecting the higher-order effect of quark feedback to
gluons within the gluon jet, the attenuation is described in moment
space\footnote{The moments of the function $f(z)$ are defined by $f(m) =
  \int_0^1 dz z^{m-2} f(z)$.} by the coefficient
\begin{equation}
g(m, E^2) = \left[ \frac{\alpha_s(E^2)}{\alpha_s(E_0^2)} 
\right]^{2\gamma_m/\beta_0} 
\end{equation}
for the evolution from the energy $E_0$ to $E$. $\gamma_m$ is the
anomalous dimension related to the gluon splittings $g \rightarrow
gg,~q\bar{q}$:
\begin{equation}
\gamma_m = \frac{3}{2} \left( - \frac{1}{3} - \frac{2N_F}{9} +
  \frac{4}{m(m-1)} + \frac{4}{(m+1)(m+2)} - 4 \sum_{j=2}^{n} \frac{1}{j}
\right) 
\end{equation}
with $\beta_0 = 11 - 2N_F/3$ where we take $N_F = 4$ active light
flavours.

\noindent
\underline{Summary:} The final gluon-to-gluonium fragmentation function
$D_g^G$ at energy $E$ which incorporates the perturbative and
non-perturbative effects is found by convoluting the perturbative
splitting function with the non-perturbative fragmentation function; in
moment space:
\begin{equation}
D_g^G(m,E^2) = d_g^G(m) \left[ \frac{\alpha_s(E^2)}{\alpha_s(E_0^2)}
\right]^{2\gamma_m/\beta_0} 
\end{equation}
with $d_g^G$ describing the non-perturbative fragmentation process at
the energy $E_0 \simeq M_G^2/\Lambda$ as argued before. 

Transformed back to momentum space, the fragmentation function $D_g^G(z, 
E^2)$ is illustrated for the gluon energy $E = 30$ GeV by the full
curves in Fig.\ \ref{fig2}. The fragmentation function is shown for two
values of the initial energy $E_0$ for the definition of the
non-perturbative function $d_g^G(z)$. The variation illustrates the
inherent uncertainties due to the qualitative estimate of $E_0$. 

However, despite of these quantitative uncertainties the qualitative
picture emerges quite clearly: A {\it hard component} is expected to be
present when gluonia are formed in the fragmentation of gluon jets.
Gluonium particles will also be generated by gluon-gluon fusion
mechanisms at low energies in the plateau region of the gluon jets.
These additional mechanisms will increase the overall multiplicity of
gluonia in the jet, yet they will not reduce the particle yield
generated by the hard component in gluon fragmentation to gluonium at
large $z$. The hard component may experimentally be quite helpful for
the search of these novel particles since the reconstruction is easier
in a phase space region of low hadronic population.

\vspace{5mm}

\noindent
\underline{Acknowledgement:} We thank P.\ Roy for an inspiring
discussion on the search for gluonia in gluon fragmentation. 


\vspace{5mm}

\noindent
\underline{Note:} After writing this letter, we received the paper of
Ref.\ \cite{ochs} in which elements of the production rate for gluonia
in gluon fragmentation have been discussed.



\begin{thebibliography}{99}

\bibitem{fritzsch}
H.\ Fritzsch and M.\ Gell-Mann, Proceedings, {\it XVI Int.\ Conference
  on High Energy Physics}, FNAL 1972; 
H.\ Fritzsch and P.\ Minkowski, Nuovo Cim.\ {\bf 30A} (1975) 393.

\bibitem{chan} 
M.\ Chanowitz and S.\ Sharpe, Nucl.\ Phys.\ {\bf B222} (1983) 211;
T.\ Schafer and E.\ V.\ Shuryak, Phys.\ Rev.\ Lett.\ {\bf 75} (1995)
1707; 
G.\ S.\ Bali et al., (UKQCD Collaboration), Phys.\ Lett.\ {\bf B309}
(1993) 378;  
D.\ Weingarten, Nucl.\ Phys.\ B (Proc.\ Suppl.)\ {\bf 34} (1994) 29.

\bibitem{roy+walsh}
P.\ Roy and T.\ F.\ Walsh, Phys.\ Lett.\ {\bf B78} (1978) 62.

\bibitem{close}
F.\ E.\ Close and A.\ Kirk, Phys.\ Lett.\ {\bf B397} (1997) 333; ibid.\
{\bf B410} (1997) 353 (Erratum).

\bibitem{landua}
Particle Data Group, C.\ Caso et al., Eur.\ Phys.\ J.\ {\bf C3} (1998)
1;  
M.\ E.\ Pennington, Proceedings, {\it Workshop on Photon Interactions
  and the Photon Structure}, ed.\ G.\ Jarlskog and T.\ Sjostrand, Lund
1998, [hep-ph/9811276].

\bibitem{roy} 
P.\ Roy and K.\ Sridhar, JHEP 9907 (1999) 13.

\bibitem{peterson}
C.\ Peterson, D.\ Schlatter, I.\ Schmitt and P.\ Zerwas, Phys.\ Rev.\
{\bf D27} (1983) 105.

\bibitem{nishijima} K.\ Nishijima, {\it Fundamental Particles}, Benjamin 
  1964. 

\bibitem{ap} 
G.\ Altarelli and G.\ Parisi, Nucl.\ Phys.\ {\bf B126} (1977) 298. 

\bibitem{ochs}
P.\ Minkowski and W.\ Ochs, e-preprint hep-ph/0003125. 
\end{thebibliography}
\end{document}